\documentclass[prl,reprint]{revtex4-1}
\usepackage{graphicx}
\usepackage{epsfig}
\def\ket#1{|\,#1 \,\rangle}

\begin{document}
\date{\today}
\author{Boon Leng Chuah}
\author{Nick C. Lewty}
\author{Murray D. Barrett}
\affiliation{    Center for Quantum Technologies and Department of
  Physics, National University of Singapore, 3 Science Drive 2, 117543 Singapore}
\title{State detection using coherent Raman repumping and two-color Raman transfers}
\begin{abstract}
We demonstrate state detection based on coherent Raman repumping and a two-color Raman state transfer.  The Raman coupling during detection selectively eliminates unwanted dark states in the fluorescence cycle without compromising the immunity of the desired dark state to off-resonant scattering. We demonstrate this technique using $^{137}\mathrm{Ba}^+$ where a combination of Raman coupling and optical pumping leaves the $D_{3/2}$ $\ket{F''=3,m_F''=3}$ metastable state optically dark and immune to off-resonant scattering.  All other states are strongly coupled to the upper $P_{1/2}$ levels. We achieve a single shot state-detection efficiency of $89.6(3)\%$ in a $1\mathrm{ms}$ integration time, limited almost entirely by technical imperfections. Shelving to the $\ket{F''=3,m_F''=3}$ state before detection is performed via a two-color Raman transfer with a fidelity of $1.00(3)$.
\end{abstract}
\maketitle
State detection is an important tool in atomic physics and is used in a wide variety of experiments, most notably those associated with frequency metrology \cite{clocks} and quantum information \cite{bible}.  In its simplest form, state detection is implemented by closed cycle optical pumping \cite{Wineland1980}.  This provides a distinction between ``bright'' and ``dark'' states by their state dependent fluorescence.  Bright states are part of the optical pumping cycle and provide a high level of fluorescence.  Dark states are off resonant with or decoupled from the excitation lasers and the fluorescence is suppressed.  More generally, the dark state may be coherently transferred or shelved to a far off-resonant state to enhance its immunity to off resonant excitation \cite{Steane1,Blinov1}.  For ions with low lying D-states, such as $\mathrm{Ca}^+$, $\mathrm{Sr}^+$, or $\mathrm{Ba}^+$, the shelving step is a necessity and usually involves driving a very narrow transition by direct excitation with a narrow bandwidth laser.

In this paper we demonstrate that state detection efficiency can be enhanced by utilizing coherent Raman coupling within the detection cycle.  The Raman coupling eliminates unwanted dark states, including coherent population trapping effects \cite{Monroe2, Steane1}, without compromising the immunity of the shelved state to off-resonant optical pumping.  We also demonstrate that shelving to long lived meta-stable states can be achieved using a two color Raman transfer.  This significantly relaxes the technical requirements of the lasers compared to the those needed for direct excitation \cite{Blinov1}.

\begin{figure}[t]
  \centering
  \includegraphics{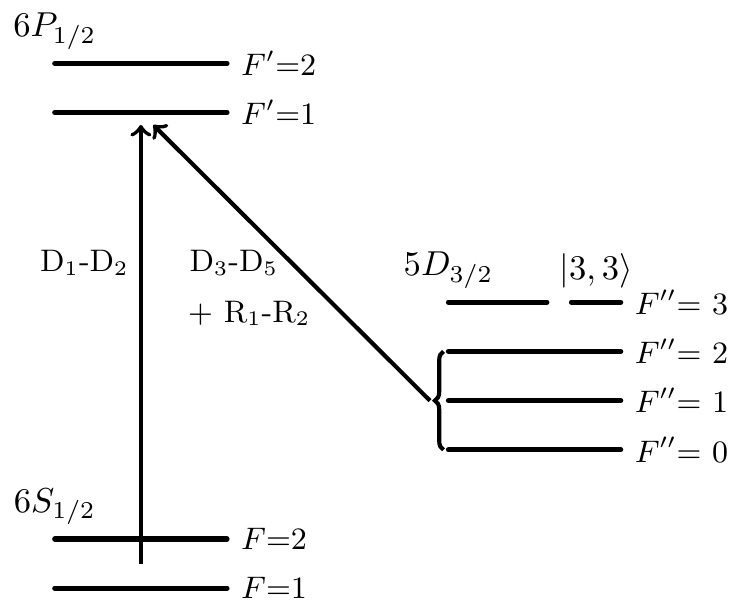}
  \caption{State detection scheme. We fluoresce on the transitions from the $\ket{F=1,2}$ ground states to the $\ket{F'=1}$ excited state via beams D1 and D2 and use the the $\ket{F''=3, m_F''=3}$  state as the dark state. By using near resonant $\pi$ polarized beams D3-D5 and coherent coupling via the Raman pair R1 and R2, we are able to repump out of the $5D_{3/2}$ meta-stable manifold without compromising the immunity of the dark state to off-resonant excitations.}
  \label{fig:simple-scheme}
\end{figure}
\begin{figure}[t]
  \centering
  \includegraphics{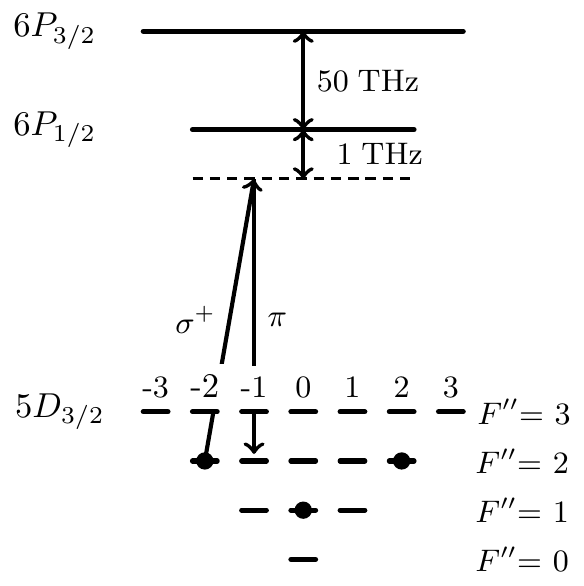}
  \caption{Coherent Raman repumping. We use $\pi$-polarized light near resonance with the $F''=0,1,2$ to $F'=1$ transitions to repump out of the $5D_{3/2}$ manifold.  This leaves three unwanted meta-stable states indicated by filled circles. To clear these we use a pair of Raman beams which are detuned by $\approx 1\,\mathrm{THz}$ from the $6P_{1/2}$ levels.  The beams are polarized $\pi$ and $\sigma^+$ and couple the $\ket{F'',m_F''}$ states to the $\ket{F'',m_F''-1}$ states via the $6P_{1/2}$ levels for all values of $F''$ and $m_F''$ with the exception of the dark state $\ket{F''=3,m_F=3}$.  Raman coupling to this state is only achieved via the $P_{3/2}$ which is a further $50\,\mathrm{THz}$ from resonance with the Raman lasers. Furthermore, since all of the detection lasers near $650\,\mathrm{nm}$ are polarized $\pi$ or $\sigma^+$, off resonant scattering out of the dark state is heavily suppressed. We also note that the Land\'{e} g-factor, $g_F$, does not depend on $F$ for the $D_{3/2}$ states.  Thus, the Zeeman splittings due to the quantizing magnetic field shifts all the Raman resonances by an equal amount and so the Raman resonances are the same for all $\ket{F'',m_F''}$ to $\ket{F'',m_F''-1}$ transitions.}
  \label{fig:raman-repumping}
\end{figure}

Consider the level scheme of $^{137}\mathrm{Ba}^+$ as shown in Fig.~\ref{fig:simple-scheme}.  The relevant levels are the $6S_{1/2}$, $6P_{1/2}$, and $5D_{3/2}$ levels, with spin quantum numbers $F$, $F'$ and $F''$ respectively.  For detection we couple the $F=1,2$ and $F''=0,1,2$ states to the $F'=1$ excited states.  This leaves the state $\ket{F''=3,m_F''=3}$ optically dark due to the selection rule $\Delta F=0,\pm 1$.  Using resonant optical pumping to clear the $F''=2$ levels requires the use of polarization with a $\sigma^-$ component.  This results in a significant depumping of $\ket{F''=3,m_F''=3}$ due to off-resonant coupling to the $\ket{F'=2,m_F'=2}$ state.  To avoid this, we instead use only $\pi$ polarization to drive the $F''=0,1,2$ to $F'=1$ transitions. The $\ket{F''=3,m_F''=3}$ state is then no longer coupled to any of the $P_{1/2}$ levels but results in unwanted dark states, namely $\ket{F''=2,m_F''=\pm 2}$ and $\ket{F''=1,m_F''=0}$. These states are coupled back into the detection cycle via coherent Raman coupling as illustrated in Fig.~\ref{fig:raman-repumping}.

% \begin{figure}
% \includegraphics[width=3.0in]{repumperscheme.eps}
% \caption{\label{Detection}Energy level diagram illustrating the relevant levels and laser couplings used. D1 and D2 are polarized $\sigma^+$ and $\sigma^-$ respectively. An EOM provides sidebands separated by the $8.036\,\mathrm{GHz}$ allowing coupling of both ground state hyperfine levels to the $P_{1/2}$, $F^\prime=1$ levels.  Further modulation at $1.488\,\mathrm{GHz}$ allows addressing of the $P_{1/2}$, $F^\prime=2$ levels. Lasers D3-D5 are $\pi$ polarized, D6 is linear polarized and perpendicular to the quantizing magnetic field. R1 and R2 are detuned $\approx 1\,\mathrm{THz}$ below the $P_{1/2}$ states and couple $\ket{F^{\prime\prime},m_F^{\prime\prime}}$ to $\ket{F^{\prime\prime},m_F^{\prime\prime}-1}$ for all $m_F^{\prime\prime}\neq 3$ via a two photon Raman process.  Similarly R3 and R4 provide a Raman coupling between $\ket{F^{\prime\prime}=3,m_F^{\prime\prime}=3}$ and $\ket{F=2,m_F=2}$}.
% \end{figure}

Details of the lasers involved are as follows.  The $S_{1/2}$ hyperfine levels are coupled to the $F'=1$ levels with two cooling lasers at $493\,\mathrm{nm}$.  These two beams are denoted D1 and D2 and are polarized $\sigma^+$ and $\sigma^-$ respectively. In order to address both hyperfine levels of the $S_{1/2}$ manifold, an electro-optic modulator (EOM) is used to provide sidebands separated by the $8.036\,\mathrm{GHz}$ hyperfine splitting.  Repumping from the $D_{3/2}$ states is achieved using three $\pi$ polarized laser beams, denoted D3-D5, that are near resonant with the $F''=0,1,2$ to $F'=1$ transitions respectively, and the Raman pair R1 and R2 as detailed in Fig.~\ref{fig:raman-repumping}.  An additional laser D6 tuned to the $F''=3$ to $F'=2$ transition and linear polarized at right angle to the B-field facilitates emptying of the $F''=3$ levels during cooling.

The experiments are carried out in four-rod linear Paul trap similar to that described in \cite{Berkeland1,Chapman1}.  The trap consists of
four stainless steel rods of diameter $1.2\,\mathrm{mm}$ whose centers are arranged on the vertices of a square with a side of length $3.6\,\mathrm{mm}$.  A $3.7\,\mathrm{MHz}$ rf potential with an amplitude of $500\,\mathrm{V}$ is applied via a step-up transformer to two diagonally opposing electrodes.  A small dc voltage applied to the other two electrodes ensures a splitting of the transverse trapping frequencies and rotates the principle axes of the trap with respect to the propagation direction of the cooling lasers.  Axial confinement is provided by two axial electrodes separated by $5\,\mathrm{mm}$ and held at $3\,\mathrm{V}$.  Using this configuration we achieve trapping frequencies of ($\omega_x$, $\omega_y$, $\omega_z$)/2$\pi$ $\approx$ (388, 348, 90) kHz.

\begin{figure}
\includegraphics[width=3.5in]{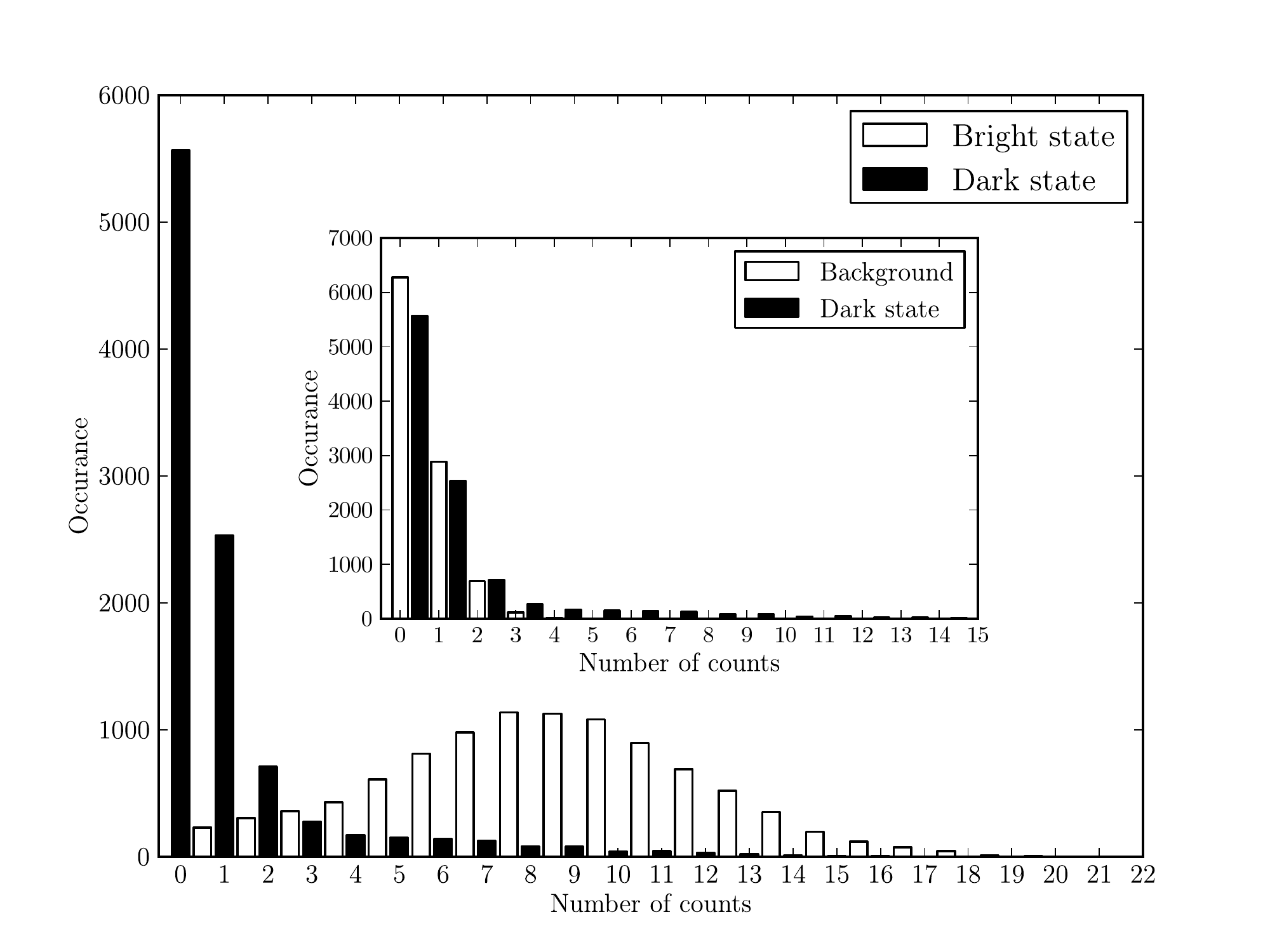}
\caption{\label{Histograms}Histogram of photon counts for two cases: one in which the ion is first optically pumped to $\ket{F''=3,m_F''=3}$ (dark) and one in which the ion is optically pumped to $S_{1/2}$ levels (bright). Each distribution corresponds to $10000$ detection events each with a $1\,\mathrm{ms}$ duration. The inset shows the comparison between the dark state and background distributions with the background being obtained with no ion present.}
\end{figure}

To assess the detection efficiency, we first use optical pumping in order to prepare the bright and dark states.  The dark state is prepared using beams D1-D5, R1 and R2. An additional frequency modulation of $1.488\,\mathrm{GHz}$ applied to D1 and D2 provides efficient optical pumping into the $F''=3$ levels via the upper $F'=2$ levels.  Since the $\ket{F''=3,m_F''<3}$ states are only emptied by off-resonant excitation from laser beams D3-D5, we use an extended optical pumping time of $1\,\mathrm{ms}$.  For the bright state, any of the $S_{1/2}$ states can be used. Thus it is sufficient to simply empty the $D$ states which is achieved using D3-D6, R1 and R2. In Fig.~\ref{Histograms} we give the histogram of photon counts for both the bright ($S_{1/2}$) and dark ($\ket{F''=3,m_F''=3}$) states. Each distribution corresponds to $10000$ detection events each with a $1\,\mathrm{ms}$ duration.  From these histograms we obtain a detection efficiency of $89.6(3)\%$ as defined in \cite{lucas1}.  By comparison, if we use direct excitation of the $\ket{F''=2,m_F''=2}$ states by rotating the polarization of D5 by $90\deg$, the dark state depumps with an estimated timescale of $\approx 100\mu s$.  In this case the fluorescence histograms are almost identical giving a detection efficiency of approximately $50\%$.

The non-Poissonian nature of the histograms is indicative of both technical imperfections and limits of the detection scheme due to off-resonant processes.  These effects result in optical pumping out of the bright and dark states during the detection time. As shown in \cite{Monroe1} the distributions can be modeled by
\begin{eqnarray*}
P_\mathrm{b}(n) & = & e^{-\beta_b (\bar{n}_b-\bar{n}_d)}\frac{e^{-\bar{n}_b}\bar{n}_b^n}{n!}+\frac{\beta_b e^{\beta_b \bar{n}_d}}{(1+\beta_b)^{n+1}} \\
& & \times \big\{\mathcal{P}(n+1,(1+\beta_b)\bar{n}_b)-\mathcal{P}(n+1,(1+\beta_b)\bar{n}_d)\big\}
\end{eqnarray*}
for the bright state and
\begin{eqnarray*}
P_\mathrm{d}(n) & = & e^{-\beta_d (\bar{n}_b-\bar{n}_d)}\frac{e^{-\bar{n}_d}\bar{n}_d^n}{n!}+\frac{\beta_d e^{-\beta_d \bar{n}_b}}{(1-\beta_d)^{n+1}}  \\
& & \times\big\{\mathcal{P}(n+1,(1-\beta_d)\bar{n}_b)-\mathcal{P}(n+1,(1-\beta_d)\bar{n}_d)\big\}
\end{eqnarray*}
for the dark state where $\mathcal{P}(n+1,x)=\frac{1}{n!}\int_0^x t^n e^{-t}\,\mathrm{d}t$ is the lower regularized incomplete gamma function.  The parameter $\bar{n}_d$ ($\bar{n}_b$) characterize the mean number of photons collected for the dark (bright) state.  The parameter $\beta_d$ ($\beta_b$) is related to the timescale $\tau_d$ ($\tau_b$) for which the ion is optically pumped out of the dark (bright) state via the expression
\[
\beta_k = \frac{\tau_D}{\tau_k(\bar{n}_b-\bar{n}_d)}
\]
where $\tau_D$ is the detection time.  Fitting to these distributions we find parameters $\bar{n}_d=0.44(2)$, $\bar{n}_b=8.7(1)$, $\tau_d=5.8(4)\,\mathrm{ms}$, and $\tau_b=3.4(2)\,\mathrm{ms}$ with a reduced $\chi^2$ of 0.91. The errors indicated are the $99\%$ confidence intervals estimated from the fit.

The parameters $n_d$ and $n_b$ can be independently assessed. The background distribution shown in the inset of Fig.~\ref{Histograms} has a measured $\bar{n}=0.47(2)$ in experimental agreement with $\bar{n}_d=0.44(2)$.  Turning on D6 during detection of the bright state eliminates depumping of this state. In this case we measure a distribution well characterized by a Poisson distribution with mean photon number $\bar{n}=8.5(1)$ consistent with $\bar{n}_b=8.7(1)$.  The pumping times $\tau_d$ and $\tau_b$ cannot be readily measured but their values may be roughly estimated as we now discuss.

Off resonant pumping of the bright state to the dark state skews the photon distribution towards lower counts. We note that it is off-resonant pumping into the dark state that is important and not just scattering into the $F''=3$ level.  The reason for this is that scattering into $\ket{F''=3,m_F''<3}$ states is offset by further off-resonant scattering back into the detection cycle.  The D5 beam, which is resonant with the $F''=0$ to $F^{\prime}=1$ transition, is only $394\,\mathrm{MHz}$ detuned from the $F''=3$ to $F'=2$ transition. Thus the $\ket{F''=3,m_F''<3}$ states are returned to the detection cycle in only a few tens of microseconds.  On the other hand, scattering into the dark state can only occur via the $\ket{F^{\prime}=2,m_F=2}$ state.  The only states coupled to this level are the $\ket{F''=2,m_F=2}$ state, where the relevant detunings are greater than $850\,\mathrm{MHz}$, and two $S_{1/2}$ states where the relevant detunings are $\approx 1.5\,\mathrm{GHz}$.  Moreover, only about 1 in 8 decays from $\ket{F'=2,m_F=2}$ result in population of $\ket{F''=3,m_F=3}$.  Assuming an equal distribution across the populated ground states we estimate the time scale for decay into the $\ket{F''=3,m_F=3}$ state to be $\approx 9\,\mathrm{ms}$ \cite{London}.  Although this is significantly larger than the timescale inferred from the fit, this estimate is dependent on the assumed distribution across the ground states \cite{London}.

Photon counts of $n>2$ in the dark state distribution indicate depumping of $\ket{F''=3,m_F=3}$ state within the detection time.  Since all of the $650\,\mathrm{nm}$ detection lasers are $\pi$ or $\sigma^+$ polarized this depumping can only occur due to coupling to the $P_{3/2}$ levels, imperfect polarization or stray magnetic fields. Coupling to the $P_{3/2}$ levels results in either direct spontaneous emission out of the dark state or an unwanted Raman coupling between the $\ket{F''=3,m_F=3}$ and $\ket{F''=3,m_F=2}$ states and subsequent off-resonant excitation back into the scattering cycle.  However, the fine structure splitting in Barium is $50\,\mathrm{THz}$ and the relevant dipole matrix elements are small.  For our beam parameters we estimate spontaneous emission rates of less than 1 photon/s.  For the unwanted Raman resonance, we also estimate a Rabi rate of less than $5\,\mathrm{kHz}$. This is more than two orders of magnitude smaller than the Rabi rates between other Zeeman sublevels.  Since the width of the Raman transition is essentially given by the Rabi rate, we can easily detune from the undesired resonance without significantly influencing the others.  To test this we have changed the frequency difference of the Raman beams by up to $1\,\mathrm{MHz}$ with no discernable change to the detection efficiency. This suggests that the depumping evident in Fig.~\ref{Histograms} is due to imperfect polarizations of the detection beams or stray magnetic fields.  However, if we assume a polarization extinction ratio of $30\,\mathrm{dB}$ for the resonant $\pi$ beams, we calculate the time scale for depumping of the dark state to be $\approx 25\,\mathrm{ms}$.  This is significantly larger than that indicated by the fitting parameter $\tau_d$ so other depumping mechanisms may be present.

Since the non-zero value of $\bar{n}$ for the dark state distribution arises solely due to technical limitations, we can directly infer a lower bound for the maximum detection efficiency based on the measured histograms.  If the technical issues could be partially or completely mitigated, we could reduce the detection error to the probability that the bright state renders zero photon counts.  This is currently given by $2.3(2)\%$ and so detection efficiencies above $97\%$ should be achievable.  Further improvements may be possible through additional optimization of laser parameters and improvements in the imaging system.

\begin{figure}
\includegraphics[width=3.5in]{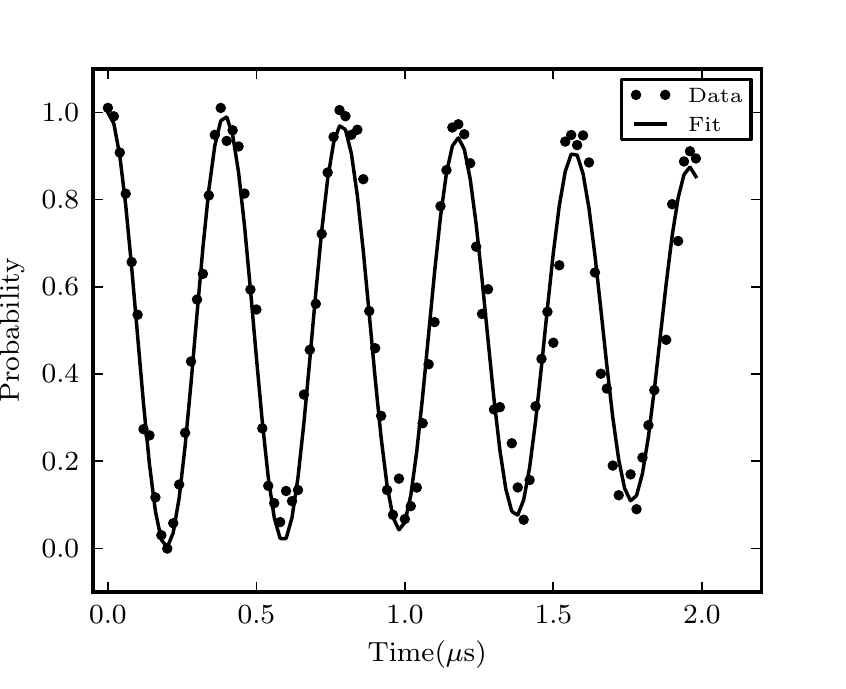}
\caption{\label{Shelving}Rabi oscillation induced by two photon Raman transition between the $S_{1/2}$ $\ket{F=2,m_F=2}$ state and the $D_{3/2}$ $\ket{F''=3,m_F=3}$ state. Each point represents $1000$ detection events and the probability of being in the $\ket{F=2,m_F=2}$ state is based on a fit to the photon distributions using dark and bright state reference histograms as a basis.  The solid curve is a fit assuming a thermal distribution along the axial direction of the trap. Fitting parameters are the mean vibrational number, $\bar{n}$, the Raman Rabi rate, $\Omega_R$. We find $\bar{n} = 14(2)$, and $\Omega=2\pi\times 2.60(1)\,\mathrm{MHz}$. The Lamb-Dicke parameter is determined from the measured trap frequency of $90\,\mathrm{kHz}$ and the beam geometry giving $\eta=0.044$.}
\end{figure}

To achieve state discrimination between $S_{1/2}$ ground states, we require a state transfer between the $\ket{F=2,m_F=2}$ ground state and the $\ket{F''=3,m_F''=3}$ state.  In principle this transition could be driven directly by a single laser \cite{Blinov1}.  However, this is a very narrow quadrupole transition and the implementation is technically demanding. Instead, we use a two photon Raman transition using lasers at $493\,\mathrm{nm}$ and $650\,\mathrm{nm}$ tuned approximately $1\,\mathrm{THz}$ below the $P_{1/2}$ states.  Both lasers are locked to the same reference cavity which provides individual linewidths below $10\,\mathrm{kHz}$ and a drift in the relative frequency of $\approx 200\,\mathrm{kHz/hour}$.  This is sufficient to ensure that the laser stability does not contribute to the dephasing \cite{Steane2}.  The resulting Rabi oscillation is illustrated in Fig.~\ref{Shelving}.  Full state transfer is achieved in a little over $200\,\mathrm{ns}$ and the fitted histogram at this point gives a transfer probability of $1.00(3)$.  The residual dephasing apparent in Fig.~\ref{Shelving} is most likely due to motional coupling.  Even though the lasers are co-propagating, the large difference in wavelength still results in a significant coupling to the motion. Assuming a thermal distribution of vibrational states and neglecting transitions between them, the Rabi oscillation is described by \cite{bible}
\[
P_\mathrm{bright}=\sum_{n=0}^\infty \frac{\bar{n}^n}{(1+\bar{n})^{n+1}} \cos^2\left(\frac{1}{2}\Omega_{n,n} t\right)
\]
where $\Omega_{n,n}$ is the Rabi rate for the $n^\mathrm{th}$ vibrational state.  In Fig.~\ref{Shelving} we include a fit to the data based on this expression using two fitting parameters, $\bar{n}$ and $\Omega_R$ where $\Omega_R$ sets the overall scale of the $\Omega_{n,n}$. We only include the thermal motion along the trap axis since this is the only motion that is expected to significantly contribute to the decay.  We find good agreement for a mean motional quanta of $\bar{n}=14(2)$ which is about $7$ times lower than the expected Doppler limit given by $\bar{n}=\Gamma/(2\omega_T)$ \cite{Wineland1979}.  It is not clear why the inferred $\bar{n}$ is so far below the Doppler temperature.  In principle the Raman coupling we have employed for repumping couples to the vibrational motion, but the associated Rabi rates exceed the axial trap frequency and thus do not resolve the motional sidebands.  In any case the low $\bar{n}$ allows a very high fidelity transfer that does not limit the detection efficiency.

In conclusion, we have demonstrated a state detection scheme that utilizes Raman coupling within the florescence cycle and high fidelity state transfers using two color Raman transitions. In the case of Barium, this provides us with a practical detection scheme that can be readily used in future experiments. However the technique may be readily applicable to other ions.  In $\mathrm{Yb}^+$, for example, detection efficiency is limited by off-resonant scattering and coherent population trapping \cite{Monroe1,Monroe2}.  The techniques introduced here would eliminate both of these issues. The two color Raman transition would also be applicable to all ions with low lying D states and can also be applied to the manipulation of cold molecules \cite{molecules}.  Beyond that, the techniques demonstrated here may find wider applicability and offer greater flexibility in the choice of atom or ion for atomic physics experiments in general.
\begin{acknowledgements}
We would like to acknowledge technical assistance from Tan Ting Rei, Chan Zhanjiang, Tan Kok Chuan Bobby and Sarah Thaler and thank Markus Baden for help with preparing the manuscript.  This research was supported by the National Research Foundation and the Ministry of Education of Singapore.
\end{acknowledgements}
\end{document}